\documentclass[useAMS,usenatbib]{emulateapj}

\newcommand\msun {M$_{\odot}$}
\def\approxgt{\ifmmode \rlap{$>$}{}_{{}_{{}_{\textstyle\sim}}} \else%
$\rlap{$>$}{}_{{}_{{}_{\textstyle\sim}}}$\fi} 
\def\approxlt{\ifmmode \rlap{$<$}{}_{{}_{{}_{\textstyle\sim}}} \else%
$\rlap{$<$}{}_{{}_{{}_{\textstyle\sim}}}$\fi}

\def\arcsec{\hbox{$^{\prime\prime}$}}

\def\flx{erg cm$^{-2}$ s$^{-1}$}
\def\lum{erg s$^{-1}$}
\def\xmm{{\it XMM--Newton}}

\def\src{XRT~000519}
\def\chan{{\it Chandra}}
\def\swift{{\it Swift}}

\shorttitle{A new kind of explosive X-ray transient}
\shortauthors{Jonker et al.}

\begin{document}
\title{Discovery of a new kind of explosive X-ray transient near M~86}

\author{P.G.~Jonker \altaffilmark{1,2,3}} 
\email{p.jonker@sron.nl}

\author{A.~Glennie \altaffilmark4} 
\author{M.~Heida \altaffilmark{1,2}}
\author{T.~Maccarone \altaffilmark{5}} 
\author{S.~Hodgkin \altaffilmark{6}} 
\author{G.~Nelemans \altaffilmark{2,7}}
\author{J.C.A.~Miller-Jones\altaffilmark{8}} 
\author{M.A.P.~Torres\altaffilmark{1}} 
\author{R.~Fender \altaffilmark{4}}

\altaffiltext{1}{SRON, Netherlands Institute for Space Research, Sorbonnelaan 2,
  3584~CA, Utrecht, The Netherlands}
\altaffiltext{2}{Department of Astrophysics/IMAPP, Radboud University Nijmegen,
P.O.~Box 9010, 6500 GL, Nijmegen, The Netherlands}
\altaffiltext{3}{Harvard--Smithsonian  Center for Astrophysics, 60 Garden Street, Cambridge, MA~02138, U.S.A.}

\altaffiltext{4}{School of Physics and Astronomy, University of Southampton, Southampton, SO17 1BJ, U.K.}
\altaffiltext{5}{Department of Physics, Texas Tech University, Box 41051, Lubbock, TX 79409-1051, U.S.A.}
\altaffiltext{6}{Institute of Astronomy, Madingley Road, Cambridge CB3 0HA, U.K.}
\altaffiltext{7}{Institute for Astronomy, KU Leuven, Celestijnenlaan 200D, 3001 Leuven, Belgium}
\altaffiltext{8}{International Centre for Radio Astronomy Research, Curtin University, GPO Box U1987, Perth, WA 6845, Australia}

\begin{abstract} We present the discovery of a new type of explosive
  X-ray flash in \chan\, images of the old elliptical galaxy
  M~86. This unique event is characterised by the peak luminosity of
  $6\times 10^{42}$ \lum\, for the distance of M~86, the presence of
  precursor events, the timescale between the precursors and the main
  event ($\sim$4,000~s), the absence of detectable hard X-ray and
  $\gamma$-ray emission, the total duration of the event and the
  detection of a faint associated optical signal. The transient is
  located close to M~86 in the Virgo cluster at the location where gas
  and stars are seen protruding from the galaxy probably due to an
  ongoing wet minor merger. We discuss the possible mechanisms for the
  transient and we conclude that the X-ray flash could have been caused
  by the disruption of a compact white dwarf star by a $\sim 10^4$
  M$_\odot$ black hole. Alternative scenarios such that of a
  foreground neutron star accreting an asteroid or the detection of an
  off-axis (short) $\gamma$--ray burst cannot be
  excluded at present.

\end{abstract}


\section{Introduction}

The Universe is not static: optical, infrared, radio, X-ray and  $\gamma$-ray
observations reveal a rich diversity in variability and explosions. In the X-ray
band the observed variability on time scales of hours to days has been
attributed to exploding massive stars (\citealt{2006RPPh...69.2259M}), or the
accretion of material onto neutron stars, stellar-mass
(\citealt{2012Sci...337..540F}) or super-massive black holes such as in active
galactic nuclei (AGN; \citealt{2012ARA&A..50..455F}).  The disruption of a star by a
super-massive black hole has also been observed (\citealt{2004ApJ...603L..17K}).

While stellar black holes up to $\sim 16M_{\odot}$ (for example in M33
X-7; \citealt{2007Natur.449..872O}) and supermassive black holes in
AGN with masses of $>10^5M_{\odot}$ (\citealt{2007ApJ...670...92G})
have been identified, black holes with masses of several hundred to a
few thousand solar masses remain elusive.  Ultra-luminous X-ray
sources (ULXs) could harbor such black holes with masses in--between
the stellar--mass black holes found in X--ray binaries and the
super--massive black holes (SMBHs; $\approxgt 1\times 10^5$ \msun)
found in the centers of galaxies. The ULX near ESO~243-49 is possibly
the best intermediate-mass black hole (IMBH) candidate
(\citealt{2009Natur.460...73F}).

Theoretically, the mass of a stellar--mass black hole, formed from the
evolution of a massive star, depends on the initial mass of the
progenitor, on the supernova explosion mechanism
(\citealt{2010ApJ...714.1217B}; \citealt{2012ApJ...749...91F}) and on
how much mass is lost during the progenitor's evolution which, in
turn, is a function of the metallicity of the black hole progenitor
star. Mass is lost through stellar winds, the mass--loss rate strongly
depends on the metallicity of the star. For a low--metallicity star
($\sim0.01$ solar metallicity) it is possible to leave a black hole of
$ \approxlt 70 M_{\odot}$ (\citealt{2010ApJ...714.1217B}).  Thus,
theories do allow for more massive stellar--mass black holes than have
been found so far in our Galaxy (e.g.~\citealt{2010ApJ...725.1918O}).

In very massive stars ($\gtrsim\ 130M_{\odot}$) production of free electrons and
positrons, due to increased $\gamma$-ray production, reduce thermal pressure inside
the core. This eventually leads to a runaway thermonuclear explosion that
completely disrupts the star without leaving a black hole, causing the upper limit
for a stellar black hole of $\sim100\ M_{\odot}$. It has been suggested that
metal--free Population III stars could have had masses above this pair--instability
limit and collapsed into intermediate--mass black holes 
(\citealt{2001ApJ...551L..27M}). It has also been suggested that IMBHs may form in
the centers of dense stellar clusters via the merger of stellar mass black holes
(e.g., \citealt{2002MNRAS.330..232C}), or from the collapse of merged supermassive
stars in very dense star clusters (e.g., \citealt{2002ApJ...576..899P}).  These
massive black holes could allow for the assembly of supermassive black holes early
in the Universe (e.g.~\citealt{2010Natur.466.1049V};
\citealt{2012Sci...337..544V}).

Stellar dynamical models predict that once every $10^3$--$10^5$ year a
star in a galaxy will pass within the tidal disruption radius of the
central black hole and thus will be torn apart by tidal forces
(\citealt{2004ApJ...600..149W}). The fall-back of debris onto the
black hole produces a luminous electromagnetic flare that is
detectable in UV and X-ray light. Several UV transients coincident
with the center of a galaxy have been detected
(e.g.~\citealt{2008ApJ...676..944G}). Candidates detected so far in
X-rays using ROSAT, \xmm, and \chan\, had black body temperatures with
kT = 0.04 - 0.12 keV (\citealt{1999A&A...343..775K};
\citealt{2004ApJ...603L..17K}; \citealt{2008A&A...489..543E}).  Tidal
disruption events have a rise time of hours to weeks. The timescale
for the light curve to peak depends on the black hole mass, with lower
mass black holes having shorter rise times and on the compactness of
the disrupted star, with more compact stars, like white dwarfs, allowing
for a shorter rise time. Recently, a new manifestation of tidal
disruption events was reported. The new extreme source discovered by
\swift\, (Swift J164449.3+573451) is caused by relativistic jet
emission launched due to a tidal disruption event
(\citealt{2011Sci...333..199L}; \citealt{2011Sci...333..203B}).  The
source had an extreme X-ray luminosity ($10^{47}$\lum) that lasted for
months. A power-law like spectrum was detected in X-rays. During a
possibly super-Eddington early phase in a tidal disruption event, the
source can produce brief flares, of duration of order thousand
seconds. Whereas most tidal disruption events are associated with the
center of a galaxy, in principle one could find them also outside the
core of a galaxy (see e.g.~\citealt{2012ApJ...758...28J}).

We here present the discovery of a new transient X-ray source. The
properties of this transient event seem to be most compatible with
those predicted for the tidal disruption of a compact (white dwarf)
star by an IMBH.

\section{Observations, analysis and results}

\subsection{{\it Chandra} X-ray observation} 

We identified a new type of X-ray transient in a {\it Chandra}
observation that started on 2000 May 19 UT (universal time). We
designate the name X-ray transient (XRT)~000519 to this event.  The
transient was discovered in {\it Chandra} images with Observation
Identification number 803. The source position was covered by the S4
CCD of the ACIS-S array of CCD detectors
(\citealt{1997AAS...190.3404G}). We reprocessed and analyzed the data
using the {\sc CIAO 4.5} software developed by the \chan\, X--ray
Center and employing {\sc CALDB} version 4.5.5.1. The data telemetry
mode was set to {\it very faint}, which allows for a thorough
rejection of events caused by cosmic rays.

The ACIS-S4 CCD is known to be suffering from an error in the readout where charge
unrelated to the arrival of X-ray photons is deposited during read-out of the CCD.
These events can be removed in the subsequent processing of the event file by using
the {\sc destreak} command in the {\sc ciao} software suite provided by the {\it
Chandra} X-ray Center (\citealt{2006SPIE.6270E..60F}). \src\, is still detected at
a very high significance level after we run the {\sc destreak} command. However,
given that this tool also removes some charge from genuine X-ray photons when the
X-ray source is very bright, we extracted the data from the peak of \src\, without
applying the {\sc destreak} step.

Using the {\sc wavdetect} tool on the X-ray image allows us to locate the event at
(J2000) right ascension 12${\rm ^h}$25${\rm ^{min}}$31.64${\rm ^{s}}$ and
declination +13$^\circ$03$^{\prime}$58.8$^{\prime\prime}$ with an estimated 68 per
cent confidence uncertainty in the position of 1$^{\prime\prime}$ due to the large
angle of 13.37$^\prime$ between the optical axis of the {\it Chandra} mirrors and
the location of the source on the detector. We investigated the point-spread
function of the source by performing a {\sc marx} simulation for such an off-axis
angle and we find that the observed point-spread function is consistent with that
expected on the basis of the simulation. Due to the large off-axis angle the source
photons were spread over hundreds of detector elements, making spectral distortion
due to photon pile-up minimal.

\begin{figure*} \includegraphics[angle=0,width=8cm,clip]{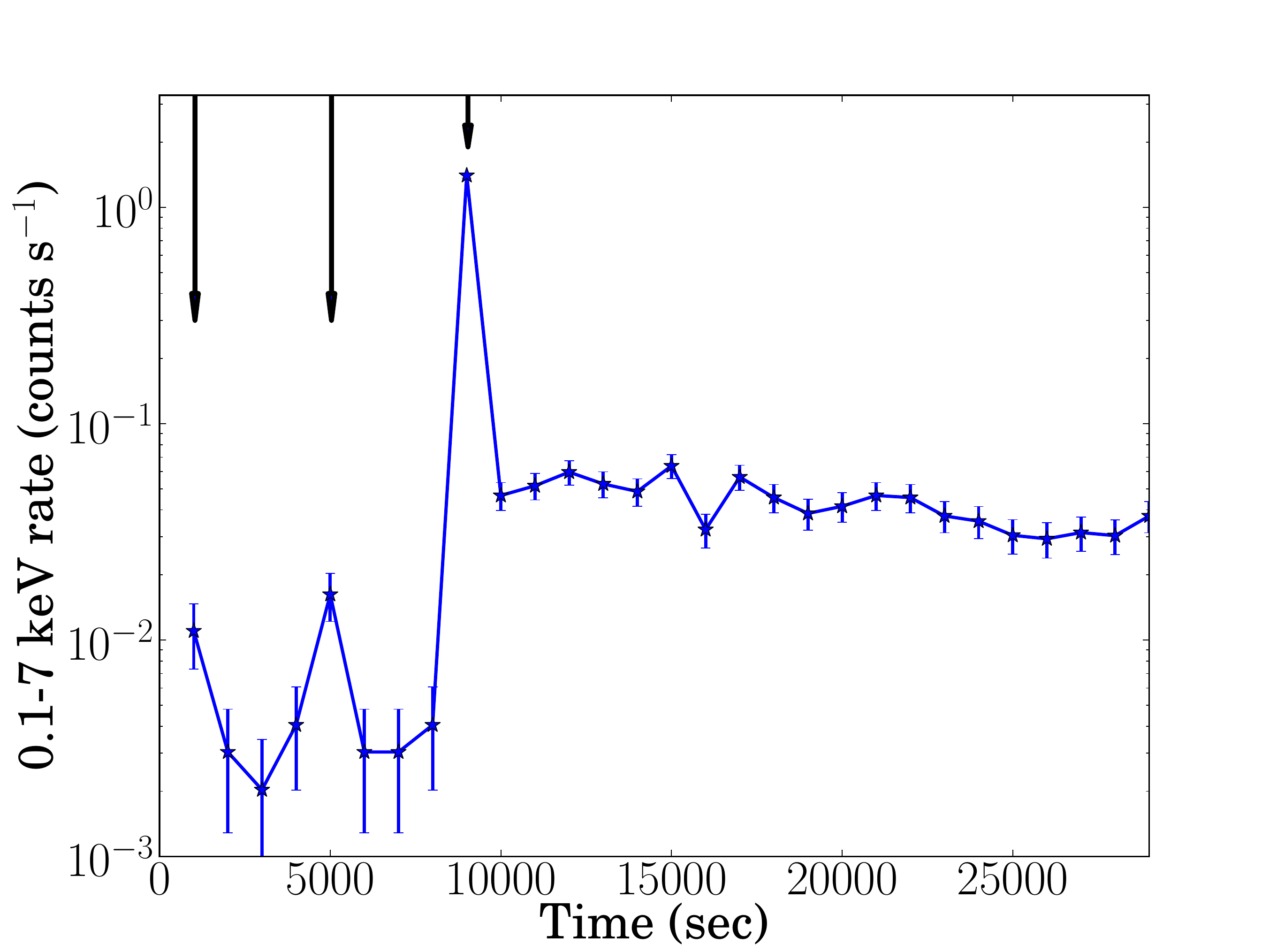}
\includegraphics[angle=0,width=8cm,clip]{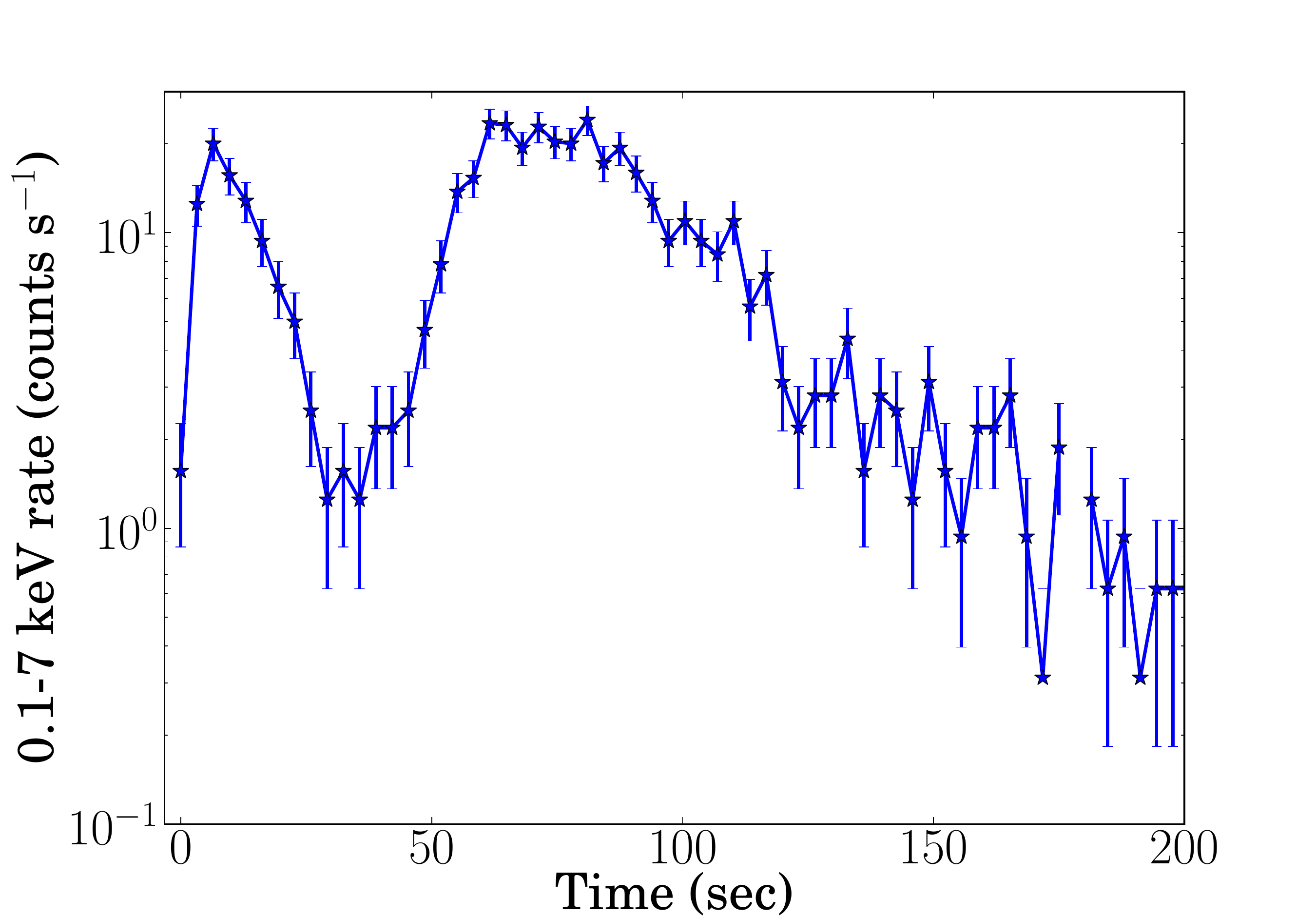}
\caption{ {\it Left panel:} The light curve of the transient X-ray
  event XRT~000519 in the 0.1-7 keV energy range at a resolution of
  1,000~s, to show the late time power-law decay and the presence of
  the precursor events, taking place approximately 4,000~s and 8,000~s
  before the main event. The precursors and the main event are
  indicated by arrows separated by 4,000~s. Time zero on the X-axis
  corresponds to the start of the {\it Chandra} observation. {\it
    Right panel:} Zoom-in on the third (main) peak in the light curve
  of the transient X-ray event XRT~000519 in the 0.1-7 keV energy
  range at the full time resolution of 3.2~s afforded by the {\it
    Chandra} detector used. The peak count rate of the second bright
  peak corresponds to a 0.1--7 keV flux of 2$\times 10^{-10}$ erg
  cm$^{-2}$ s$^{-1}$. Zero on the X-axis corresponds to the start of
  the main X-ray event at T=9,571~s in the left panel.\label{lc}}
\end{figure*}

The light curve of the main event, plotted at the maximum time
resolution of 3.24~s, is double peaked (see Figure ~\ref{lc}). The
source flux rises within 10~s from being undetected to a peak count
rate of 20 counts s$^{-1}$, it decays more slowly over a period of
20~s to a count rate just above 1 count s$^{-1}$, to rise again to the
peak count rate of 24 counts s$^{-1}$.  The second peak has a flat top
that lasts around 20~s, before gradually decreasing on a time scale of
about 100~s, followed by a slow power law decay with an index of
-0.3$\pm$0.1 that could be followed for 2$\times 10^4$~s until the end
of the observation.

When investigating the X-ray light curve of the source we found
evidence for the presence of a precursor event of 16 X-ray photons
about 4,000~s before the main flare.  In the same 1,000~s interval of
time of the precursor, we find 4 photons due to the background in
a source-free region with the same size as the source region, on the
same CCD and at a similar off-axis angle as XRT~000519. Thus, the
chance of finding 16 photons due to random variations in the
background is less than 1 in 250,000.  Interestingly, there is
evidence for another precursor event, again about 4,000~s earlier. In
that case, we find 11 photons in 1,000~s where 3.7 are found in an
off-source region over the same time; this should happen by chance in
less than 1 in 650,000 cases. Therefore, we conclude that these two
precursor events are likely to be real. The small number of events
(three) does not allow us to conclude that the 4,000~s time scale is
periodic.

\begin{figure} 
\includegraphics[angle=0,width=8cm,clip]{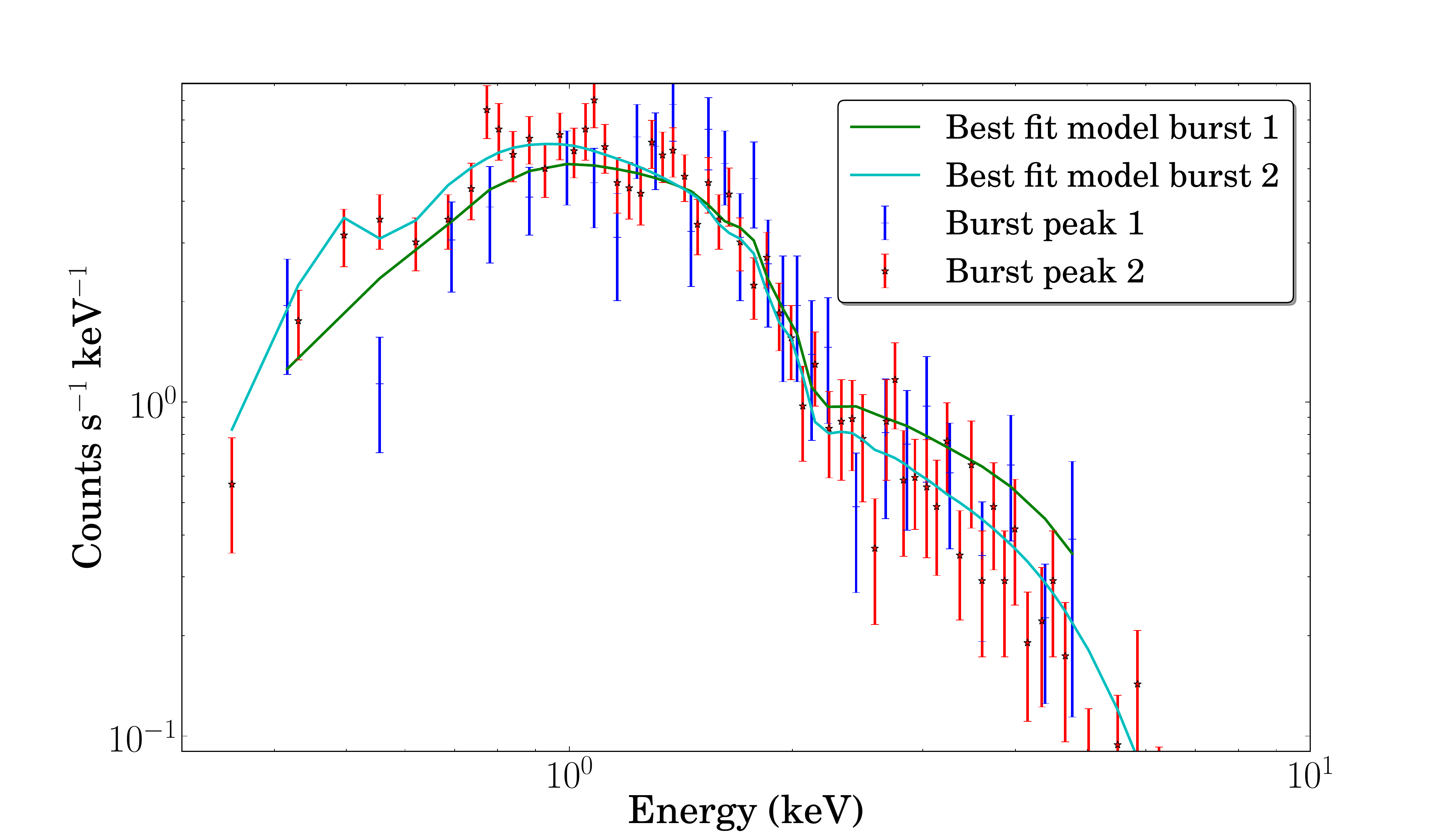}
\caption{The observed soft X-ray spectral energy
  distribution for the two large peaks of
  the transient \src. The blue data points and the best fitting
  powerlaw model (green drawn line) of the first large peak are
  plotted together with the data points (red) and the best fitting
  model (light green) for the second large peak. The second peak is
  brighter and it has a softer spectrum than the first bright peak.}
\label{xspec} \end{figure}

We extracted the source spectrum from photons in a circular region
with radius of 30\arcsec\, centered on the source position of \src\,
in the energy range of 0.1--7 keV. Background events were extracted
from an annulus centered on the position of the source with an inner
radius of 60\arcsec\, and an outer radius of 105\arcsec.  Using {\sl
  xspec} version 12.4.0ad (\citealt{ar1996}) we have fitted the
spectra of \src\ using Cash statistics (\citealt{1979ApJ...228..939C})
modified to account for the subtraction of background counts, the so
called W--statistics\footnote{see
  http://heasarc.gsfc.nasa.gov/docs/xanadu/xspec/manual/}. We have
used an absorbed power--law model ({\sl pegpwrlw} in {\sl xspec}) to
describe the data. For the extraction of the X-ray spectral parameters
we added an extinction of ${\rm N_H}=2.6\times 10^{20}$ cm$^{-2}$ due
to the Galactic foreground (\citealt{1990ARA&A..28..215D}). We did not
detect enough counts to add a component to the fit function that could
describe the influence of the potential presence of local material
causing extinction in addition to that by the Galactic foreground.
The X-ray spectrum of the first of the two bright peaks can be well
described by a powerlaw with photon index 1.6$\pm$0.1
(68\%~confidence level) where we included the effect of the Galactic
extinction in the direction of the source. The energy spectrum of the
second of the two bright peaks is softer with a power law photon index
of 1.95$\pm$0.05 (see Figure~\ref{xspec}). The fact that the second
peak has a higher peak count rate while it has a softer spectrum is in
line with the inference that the X-ray spectrum is not distorted by
the effects of photon pile-up.

We further investigated the energy associated with each of the X-ray
photons making up the precursor events. We find that most of their
photons have an energy near 1--2 keV which renders support for their
interpretation as genuine source photons rather than background
photons, as the spectrum agrees well with that found for the main
event.

\subsection{BATSE observations}

XRT~000519 was not detected by the Burst And Transient Source
Experiment (BATSE), although the source was in the BATSE field of view
at the time of the event (including the precursor
events). Extrapolating the best-fitting X-ray spectrum at peak would
have the source falling a factor of $\sim$50 below the BATSE threshold
for detection. The nearly steady emission during the second half of
the \chan\, observation falls a factor of a few below the occultation
technique's sensitivity limit.

\subsection{XMM-{\it Newton} observations of the field}

XMM-{\it Newton} X-ray observations of the region on the sky
containing the position of XRT~000519 were obtained on 2002 July 1,
2004 December 27 and 2011 June 1 (the observation identification
numbers for these observations are 0108260201, 0210270201, and
0673310101, respectively). Using the 2012 June 21 release of {\sc
  sas}, we cleaned the event lists for periods of enhanced background
leaving 78 and 79 ksec for the MOS1 and MOS2 detectors in the 2002
observation, respectively.  The source region falls off the pn CCD
during the 2002 observation.  The 2004 observation yielded 18 ksec of
cleaned pn exposure and 21.7 ksec for both MOS1 and MOS2. Finally, the
2011 observation has 40 ksec of cleaned pn exposure.

The source went undetected with a (0.5-10 keV) flux limit of $\approx
4\times 10^{-14}$ \flx\, for an assumed incident power law spectrum
with index 1.7 for the first two observations and $\approx 3\times
10^{-13}$ \flx\, for the last observation, giving a luminosity upper
limit of 1$\times 10^{39}$ \lum\, for the first two and 9$\times
10^{39}$ \lum\ for the third observation, respectively if the event
took place at the distance of M~86 of 16.2 Mpc.

\subsection{Ultra-violet, optical and near-infrared observations}

We queried various data archives to search for archival ultra-violet,
optical and near-infrared images of the field of XRT~000519. 

The Galaxy Evolution Explorer satellite (GALEX) observed the field of
the transient on 2006 March 20 for 15700~s with its near-ultraviolet
camera and for 482~s with its far-ultraviolet camara. No source was
detected in either the near-ultraviolet or the far-ultra-violet image
down to a magnitude limit of 25.5 and 23.7, respectively.

We investigated a 999~s $i^\prime$-band image
($\lambda_{central}=7743$ \AA, $\Delta \lambda=1519$ \AA) from the
Isaac Newton Telescope (INT) at La Palma, Spain, obtained on 2001
March 22 (see Figure~2). We found a faint optical star at a magnitude
of $i^\prime=24.3 \pm 0.1$ magnitude. The source is located at (J2000)
right ascension 12${\rm ^h}$25${\rm ^{min}}$31.636${\rm ^s}$ and
declination +13$^\circ$03$^\prime$58.01$^{\prime\prime}$, where the
uncertainty in the position is 0.1 arcsec in both right ascension and
declination. The angular difference between the X-ray position of
\src\, and the position of this faint $i^\prime$-band source is 0.8
arcsec. This is within the 1\arcsec\, error circle of the \chan\,
X-ray position of the source.

\begin{figure} \includegraphics[angle=0,width=8cm,clip]{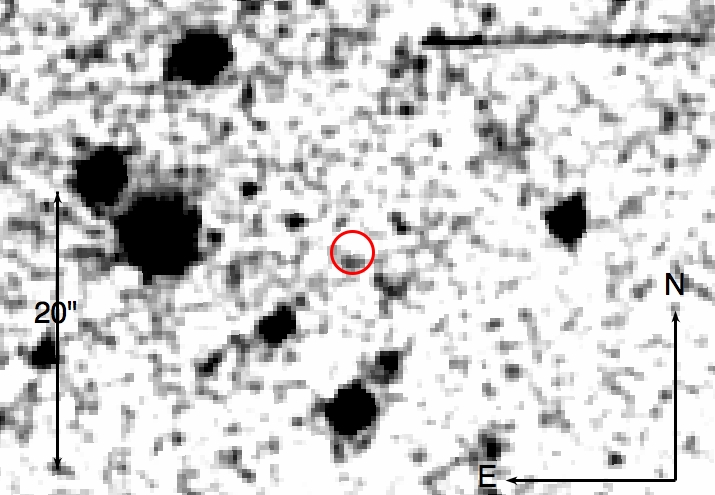}
  \caption{The i$^\prime$-band image of the field of
    \src~ obtained using the Isaac Newton Telescope on 2001 March
    22. The exposure time is 999~s and the seeing is 1 arcsec. The red
    circle indicates the position of the \chan\, X-ray event (for
    display purposes we increased the radius of the circle to 1.5
    arcsec instead of the 1 arcsec formal 68\% confidence
    uncertainty). We used a Gaussian smoothing with a kernel radius of
    2 pixels to highlight the faint source. The larger number of
    sources towards the North-East of the image is due to the presence
    of a stream of stars protruding from M~86
    (see Section 3).}
\end{figure}

\begin{figure} 
\includegraphics[angle=0,width=8cm,clip]{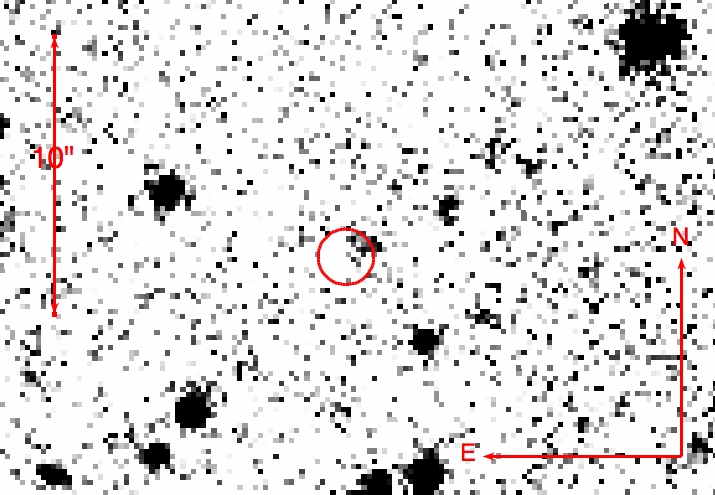}
\caption{Optical $g^\prime$-band image obtained at the
  Canada-France-Hawaii Telescope with a limiting magnitude of
  27.2. The red circle of radius 1 arcsec indicates the position and
  its 68\% uncertainty due to the location of the X-ray event
  XRT~000519. A faint point source at $g^\prime=26.8 \pm 0.1$ is found
  close to but just outside the error circle of the position of the
  transient.}
\label{optfasttrans} \end{figure}

We found archival optical images in the
$u^\prime,\,g^\prime,\,i^\prime$ and $z^\prime$-band obtained in 2009
and 2010 by the Canada-France-Hawaii Telescope (CFHT) using the
MegaPrime instrument (see Table 1). In addition, we report on
$r^\prime$ images obtained on 2005 Jan.~17 with the same
instrument--telescope combination. Using the Terapix data analysis
results, we found no source down to the limiting magnitude of the
images ($u^\prime=27.2$, $r^\prime=26$, $i^\prime=25.6$ and
$z^\prime=25.3$).  However, in the deep, median combined
$g^\prime$-band image totalling 3170~s of exposure obtained between
2009 May 22 and 25, under seeing conditions of 0.7 arcseconds, there
is evidence for the presence of a faint unresolved source at a
position of right ascension 12${\rm ^h}$25${\rm ^{min}}$31.56${\rm
  ^s}$ and declination 13$^\circ$03$^{\prime}$59.2$^{\prime\prime}$
with $g^\prime=26.8 \pm 0.1$ (see Figure~\ref{optfasttrans}). This
source position is 1.2 arcsec away from the position of \src\, and
thus falls just outside the estimated error circle. The non-detection
of the INT $i^\prime$-band source in the deep CFHT $i^\prime$-band
image obtained a few years later in time (2009 May 14-15) down to a
magnitude limit of 25.6 magnitudes, indicates that the source
magnitude decayed by at least 1.2 magnitudes, strengthening the
association of the INT optical i$^\prime$-band source with \src.

\begin{table*}
  \caption{Optical observations of the field of \src\, after its explosion
    on 2000 May 19. }
\label{tab:opt}
\begin{center}
\begin{tabular}{cccccc}
\hline
Telescope & Date & Exposure time (s) & Band & Seeing (arcsec) & Magnitude upper limit/detection \\
\hline
INT & 2001 Mar.~22 & 999 & i$^\prime$ & 1.0 & 24.3$\pm$0.1\\
SDSS & 2003 Mar.~23  & 54     & u$^\prime$ & 1.6      & $>$22   \\
SDSS & 2003 Mar.~23  & 54     & g$^\prime$  &  1.6      & $>$22.2   \\
SDSS & 2003 Mar.~23  & 54     & r$^\prime$   &  1.2     & $>$22.2  \\
SDSS & 2003 Mar.~23  & 54     & i$^\prime$    &  1.3    & $>$21.3  \\
SDSS & 2003 Mar.~23  & 54     & z$^\prime$    &  1.3    & $>$20.5  \\
CFHT & 2010 Jan.~10-21 & 6402 & u$^\prime$ & 0.88 & $>$ 27.2 \\
CFHT & 2009 May 22-25 & 3170 & g$^\prime$ & 0.69 & 26.8$\pm$0.1$^*$ \\
CFHT & 2009 May 14-15 & 2055 & i$^\prime$ & 0.54 & $>$25.6 \\
CFHT & 2010 Jan.~8-18 & 4400 & z$^\prime$ & 0.74 & $>$25.3  \\
CFHT & 2005 Jan.~17  & 1440 & r$^\prime$ & 1.0 & $>$26  \\
\end{tabular}
\end{center}
\footnotesize{$^*$ The location of this source is offset from the
  $i^\prime$-band INT detection by 1.6 arcseconds.}

\end{table*}

Sloan Digital Sky Survey images (SDSS; \citealt{2012ApJS..203...21A}) in the
$u^\prime,\,g^\prime,\,r^\prime,\,i^\prime$ and $z^\prime$-bands were obtained
closer in time to XRT~000519 than the CFHT observations, namely on 2003 March 23
(see Table~\ref{tab:opt}). In these images there is no evidence for a source in the
error circle of the {\it Chandra} X-ray image down to the limiting
magnitude of the SDSS images.

The source region was observed on 2007 May 3 by the UK infrared telescope (UKIRT)
as part of the UKIRT Infrared Deep Sky Survey (UKIDSS)
(\citealt{2007MNRAS.379.1599L}) in $Y$, $J$, $H$, and $K$, but no source was
detected in the error region of ~\src\ down to the survey limit of 18.3 magnitudes
in each of the bands.  We obtained near--infrared $K$-band follow-up observations
of the field of XRT~000519 in 2013 January using the 4.2~m William Herschel
Telescope on La Palma. In 3750~s of exposure we find no source down to a limiting
magnitude of $K>20.3$.

\subsection{Radio observations}

We also investigated archival images obtained by the VLA on 2005
Aug.~15, 2009 Aug.~3 and on 2012 May 1-2 (see Table~\ref{tab:vla}).
Although the pointing of the observations was centred on M84, the
primary beam of the VLA provides a response of 55\% at the position of
the transient (12.7 arcmin from the pointing centre).  The last, most
sensitive observation was taken with a total bandwidth of 256 MHz,
split into two 128 MHz sub-bands centred at 1452 and 1820 MHz.  The
primary calibrator was 3C~286, and the secondary calibrator was
J1254+1141.  The flux scale was set according to the coefficients
derived at the VLA by NRAO staff as implemented in the 31DEC13 version
of AIPS and CASA version 4.1.0.  Data reduction was carried out
according to standard procedures within AIPS and checked using CASA
data reduction procedures.  Images were made using Briggs
weighting with a robust value of 0, to reduce the sidelobes from M~84
(which has a peak of 130 mJy/beam) and no source was found at the
transient position down to an upper limit of 0.18 mJy/beam (three
times the root-mean-square noise of the image at the position of
\src. For the upper limits of the other two observations see
Table~\ref{tab:vla}).

\begin{table*}
  \caption{VLA L-band 1.6 GHz observations of the field of \src\, after its explosion
    on 2000 May 19. }
\label{tab:vla}
\begin{center}
\begin{tabular}{cccccc}
\hline
Program ID & Start date & Exposure time (s) & Configuration  & Flux limit (mJy) \\
\hline
AV~281 & 2005 Aug.~15 & 2490 & C & 3 \\
AM~989 & 2009 Aug.~3 & 7180 & C &  3.6 \\ 
12A-098 & 2012 May~1 \& 2 & 17770 & C $\rightarrow$CnB & 0.18 \\
\end{tabular}
\end{center}
\end{table*}

\section{Discussion}

\begin{figure*} 
\includegraphics[angle=0,width=16cm,clip]{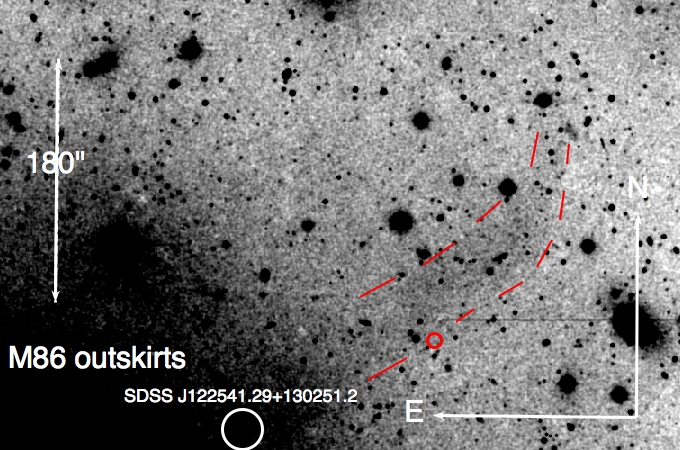}
\caption{The INT $i^\prime$-band image around the field of the
  transient (which is indicated with the red circle). The cut levels
  are chosen to highlight the presence of the stream of stars (marked
  with dashed red lines on both sides of the stream of
  stars). Potentially, the stars are stripped from the galaxy
  SDSS~J122541.29+130251.2 that in this representation is hidden in
  the glare of the stars of the outskirts of M~86 (but its position is
  shown by the white circle). SDSS spectroscopy of this galaxy shows
  that it is a starburst galaxy that could well be associated with
  M~86.}
\label{stream} \end{figure*}

We found a new transient X-ray source in an archival \chan\, observation. The source
position is 12.16 arcminutes from the centre of M~86, and it does not fall in the
M~86 $\mu_B=25$ magnitude per arcsec$^2$ isophote area
(\citealt{1991rc3..book.....D}) that has an approximate radius of 8.5 arcminutes in
the direction of the \src~ location. However, it was found that M~86 is falling
into the Virgo cluster and that gas is stripped off the galaxy
(\citealt{2008ApJ...688..208R}). The projected stripped gas lies close to the
position of \src. Furthermore, the deep optical INT images we investigated show
that stars, possibly stripped off the galaxy SDSS~J122541.29+130251.2, follow the
stripped gas, which suggests that M~86 shows signs of a recent minor merger (see
Figure~\ref{stream}). The small projected distance on the sky between the position
of \src~ and that of the stream of stripped stars strengthens the association
between \src\, and M~86. The stream of stars falls between the wedge-shaped hot gas
structure visible in the X-ray images of the field (\citealt{2008ApJ...688..208R}),
suggesting that some or all of the gas may come from the infalling galaxy, making
this a wet minor merger.

Whereas the source location is in the direction of the early type
galaxy M~86, the source could in principle be a foreground object
located in our own Galaxy, it could indeed be located at the distance
of 16.2 Mpc of M~86, or it could be located further away, well behind
M~86. As the close proximity on the sky makes a scenario where the
distance is consistent with that of M~86 appealing, we discuss here a
scenario for \src\, that is borne out by the observations, where
the source is associated with M~86. Below we discuss other scenarios
with the source placed at the distance of M~86 or at other distances.

\subsection{A transient near M~86?}
If the transient event took place at the distance of M~86 the peak
luminosity would be $6\times 10^{42}$\lum. The late time, slowly
decaying, (0.5-10 keV) flux level is 4$\times 10^{-13}$\flx, which for
the distance of M~86 would be $10^{40}$\lum. If we interpret
the peak flux as the Eddington luminosity we derive a mass estimate of
4.6$\times 10^4$ M$_\odot$.

The time scale and luminosity of the event, and the occurence in an
old population, limit the potential interpretations. We only know a
few classes of events that can release so much energy over a short
timescale. We favor the scenario where XRT~000519 is due to a tidal
disruption event (TDE) of a star by an intermediate-mass black hole
near M~86. The (quasi-) periodic recurrence time of the precursors of
$\approx$4,000~s is potentially related to the orbital period of the
stellar material present around the black hole after the tidal
disruption, with subsequent passages of a partially disrupted star
(\citealt{2013ApJ...767...25G}) or with variations in the fall-back
rate after the white dwarft has been disrupted. The latter is likely
to occur as the white dwarf will be disrupted in a strong
gravitational field regime where the orbits of the fluid elements are
not closed giving rise to the possibility of multiple shocks, however,
the current models are not yet calculated taking this into account
(see the discussion in \citealt{2009ApJ...695..404R}). This short
recurrence time of $\approx$4,000~s and the inference that the peak
flux is at or close to the Eddington luminosity, would imply a $\sim
10^4$ M$_\odot$ black hole tidally disrupting a white dwarf (see
e.g.~\citealt{2009ApJ...695..404R}; \citealt{2011MNRAS.410..359L}). In
the two observed passages prior to the main event, some of the
material of the white dwarf star is accreted by the black hole giving
off X-rays with a peak luminosity of $\sim 6\times 10^{39}$\lum. In a
subsequent orbit the self-interaction of the accreting material
increases the viscosity giving rise to the bright peaks, in line with
modelling (\citealt{1989A&A...209..103L,2009ApJ...695..404R}). It is
unlikely that the white dwarf detonates: a detonating white dwarf
would appear as a bright Type Ia supernova and the absolute
$i^\prime$-band magnitude during the INT observation is too low to be
consistent with a Type Ia supernova less than one year after the
explosion (\citealt{2007A&A...470L...1S}). The observed tail in the
\chan\, observation would be associated with the accretion of part of
the material falling back (at super-Eddington rates) towards the
intermediate-mass black hole
(\citealt{2009MNRAS.400.2070S,2011MNRAS.410..359L}). The luminosity
may well be limited to be below the Eddington limit. To explain the
two-peak structure of the main event of \src\, detailed hydrodynamical
calculations have to be performed, which is beyond the scope of this
paper.

In order for a star to be tidally disrupted by a black hole, it has to
wander close enough. The rates for this to happen are low, except for
instance in regions of high stellar density such as the centers of
galaxies or in globular clusters
(\citealt{2006MNRAS.372..467B,2009ApJ...697L..77R}). The photometry we
have is consistent with a scenario where \src\, originated in a
globular cluster, although the globular cluster would have to be at the faint end of the
globular cluster luminosity function (\citealt{1996AJ....112.1487H},
2010 edition).

Below, we discuss alternative scenarios for \src. The time scale and luminosity of
\src, and the occurence in an old population, if located at the distance of M~86,
suggests that XRT~000519 could for instance be due to two compact objects merging,
such as a neutron star - neutron star merger. Given that we detect no hard X-ray or
$\gamma$-ray emission, we would be observing the merger off-axis, unlike the
short-hard $\gamma$-ray bursts (\citealt{2011ApJ...732L...6R}). Accretion from
fall-back material onto the newly formed black hole would in this scenario account
for the late time X-ray emission. The non-detection of the source in the radio down
to 0.18 mJy on 2012 May~1$-$2 (Karl G.~Jansky Very Large Array, central frequency
1636 MHz), thus nearly 12 years after the event, only loosely constrains the
circum-stellar binary density in this scenario (for predictions for radio
emission of off-axis $\gamma$-ray bursts see \citealt{2011Natur.478...82N} and
\citealt{2011ApJ...733L..37V}).  However, the precursor events are hard to explain
in this scenario, making it less likely. An alternative, related, scenario involves
a neutron star overflowing its Roche lobe onto a black hole or another neutron
star. The neutron star mass donor will eventually drop below the minimum mass for a
neutron star.  At that point, the star should explode, and some of the debris
should be captured by the other compact object
(\citealt{1990SvA....34..595B,1993ApJ...414..717C}). The optical source could in
this scenario be interpreted as the late time fading counterpart at M${\rm
_{i^\prime}}=-6.7$ for a Distance Modulus of 31 for M~86 (which converts to an
$i^\prime$-band luminosity of 8$\times 10^{37}$\lum). However, again the precursor
events require a special explanation in this scenario.

If the source is in M~86 it could in principle be that we witnessed a
Type Ia supernova shock break-out (\citealt{2010ApJ...708.1025K}). A
problem with this scenario is that the optical supernova was not
discovered, which is unlikely given that the Virgo and M~86 region of
the sky are well monitored (\citealt{2000AJ....119.1901A}) and given
that Type Ia SNe have an absolute magnitude of -19 in the $V$-band
(\citealt{2000ARA&A..38..191H}) which means that the optical supernova
would have reached $V\sim$12, which is easily accessible also for
amateur astronomers. The detection of the faint $i^\prime$-band at
M${\rm _{i^\prime}}=-6.7$ also makes this scenario improbable as
typical type Ia supernovae do not decay that fast in the
$i^\prime$-band within a year (\citealt{2007A&A...470L...1S}).

\subsection{A foreground object?}
One could also envisage a scenario where the source is nearby, e.g.~in our own
Galaxy. For instance, the brightening in X-rays and the $i^\prime$-band could be
due to an accretion event onto an isolated, old, neutron star. The isolated neutron
star is then possibly detected in the late-time CFHT $g^\prime$-band observation.
The two detections are offset by 1.1 arcseconds in right ascension and by -1.2
arcseconds in declination giving a distance on the sky between the two positions of
1.6 arcseconds. This would yield a proper motion of 0.16 arcseconds per year with
respect to the $i^\prime$-band variable source detected in 2001. For a typical
transverse velocity of an isolated neutron star of 200 km s$^{-1}$, we derive a
distance of 250 pc for the source. At that distance the fluence of the event would
imply the accretion of around $5\times10^{15}$~g onto the neutron star. The deep
upper limits on the quiescent X-ray emission derived from the XMM-{\it Newton}
observations imply a limit on the luminosity of 3$\times 10^{29}$ \lum~ for such a
distance. This means that the quiescent neutron star luminosity is lower than what
has been observed so far for isolated neutron stars
(\citealt{2004ApJS..155..623P}), however, the current observations favor young, hot
neutron stars. Additionally, the cooler the neutron star, the more its emission
peaks towards the soft X-rays or even the ultra-violet part of the energy
spectrum.  Additional, new, optical (e.g.~Hubble Space Telescope) observations
could test this scenario by checking whether the faint $g^\prime$-band source
indeed has a proper motion of 0.16 arcseconds per year.

In an alternative version of this scenario, where we do not ascribe
the $g^\prime$-band detection to the isolated neutron star, and thus
we have no constraint on the proper motion or distance of the neutron
star, we can place it at a larger distance. This would increase the
total energy liberated in the transient event and thus the amount of
accreted mass but it alleviates the constraint on the quiescent X-ray
luminosity of the isolated neutron star. However, in both these
isolated neutron star scenarios the precursor events and their 4,000~s
time scale are difficult to understand. Furthermore, the accretion of
asteroid-sized bodies onto a neutron star are thought to produce
significant amounts of $\gamma$-ray emission
(\citealt{2011Natur.480...69C}), which is not detected from \src, implying
that for reasons unknown, the $\gamma$-ray emission is much reduced in
this case, making this scenario less likely.

\subsection{A transient (far) behind M~86?}
If we instead interpret \src\, as coming from a redshift, z, between
0.23--1.5 in line with the $g^\prime$-band  non-detection of an
unresolved faint dwarf galaxy of an absolute magnitude of $-14.5>{\rm
  M_{g^\prime}}>-19.5$ (\citealt{2005ApJ...631..208B}), then
the luminosity of XRT~000519 would be between 3$\times 10^{46}$ --
3$\times 10^{48}$ \lum. 

Then, it could potentially be interpreted as an X-ray flash such as
those found in X-rays and which are probably related to $\gamma$-ray
bursts (\citealt{2006Natur.442.1008C,2005ApJ...629..311S}).  However,
with a peak energy, ${\rm E_p}$ of about 1.5 keV, the event has a
softer X-ray spectrum than that of X-ray flashes known so far
(\citealt{2005ApJ...629..311S}).  Nevertheless, if the event was
indeed near z=1.5, thus at a luminosity of L$=3\times 10^{48}$\lum,
then the peak energy ${\rm E_p}$ of 1.5 keV and the isotropic peak
energy at the peak flux, ${\rm E_{iso}}$, are consistent with those
expected extrapolating the ${\rm E_p}$ and ${\rm E_{iso}}$ correlation
of X-ray flashes and $\gamma$-ray bursts to lower values (the
\citealt{2008MNRAS.391..577A} relation). Whereas we cannot rule out
that XRT~000519 was due to an event similar to an X-ray flash,
extending the peak energies and the isotropic luminosities for those
events a factor of several below those that have been found for this
class of flashes until now (\citealt{2005ApJ...629..311S}), the
presence of the precursor events is never seen in X-ray flashes so
far, making the association unlikely. Although, the latter could be
due to the reduced sensitivity of the satellites that detected the
X-ray flashes so far, compared with the \chan\, sensitivity. A
potential problem with this scenario and a distance of z=1.5 is that
even at 10 months after the event the absolute $i^\prime$-band
magnitude $M_{i^\prime}$ would still have to be -20.9
magnitudes. Furthermore, the projected co-location on the sky of an ongoing minor
merger event and the X-ray transient would be a chance alignment
in this scenario.

Interestingly, Shcherbakov et al.~(\citealt{2012arXiv1212.4837S}) describe
the X-ray flash reported by Campana et al. (2006) as a TDE of a white
dwarf by an intermediate-mass black hole. This shows that the
properties of these two classes of objects, X-ray flashes associated
with stellar-mass black hole formation and X-ray flashes associated
with TDEs by massive or intermediate-mass black holes, overlap. The
main reason for this is that TDEs cover X-ray luminosities ranging
from as low as
$10^{40}$\lum\,(\citealt{2008A&A...489..543E,2008ApJ...676..944G})
up to $10^{48}$ \lum\, for the Blazar-like TDE Swift~J1644+57
(\citealt{2011Sci...333..199L}). The observed peak luminosity of
XRT~000519 is in the range of distances we consider here. Thus, the
properties of XRT~000519 are also consistent with a TDE in a dwarf
galaxy at a distance between approximately 1.1 and 11 Gpc provided the
emission is strongly beamed towards us such as in Swift~J1644+57
(\citealt{2011ApJ...743..134K,2012arXiv1212.4837S}), although
again the optical $i^\prime$-band magnitude would favor distances
closer to 1.1 Gpc over one close to 11 Gpc.

\section{Conclusion}
We discovered a peculiar transient (\src) in the direction of
M~86. Furthermore, we found evidence for an ongoing wet minor merger
between M~86 and the galaxy SDSS~J122541.29+130251.2. This activity
makes it conceivable that the transient is located at the distance of
M~86. If so, its properties are consistent with a scenarion where the
transient is due to the tidal disruption of a white dwarf by an
intermediate-mass black hole. Alternative scenarios such that of a
foreground neutron star accreting an asteroid or the detection of an
off-axis (short) $\gamma$--ray burst cannot be fully excluded at
present.  Future, high resolution and deep Hubble Space Telescope
imaging should reveal the host galaxy if it was due to an event in the
background of M~86 such as a tidal disruption event at larger distance
or an off axis $\gamma$--ray burst.

\bibliographystyle{apj}

\begin{thebibliography}{}

\bibitem[\protect\citeauthoryear{{Ahn} et~al.}{{Ahn}
  et~al.}{2012}]{2012ApJS..203...21A}
{Ahn}, C.~P., et~al. 2012, \apjs, 203, 21

\bibitem[\protect\citeauthoryear{{Akerlof} et~al.}{{Akerlof}
  et~al.}{2000}]{2000AJ....119.1901A}
{Akerlof}, C., et~al. 2000, \aj, 119, 1901

\bibitem[\protect\citeauthoryear{{Amati} et~al.}{{Amati}
  et~al.}{2008}]{2008MNRAS.391..577A}
{Amati}, L., {Guidorzi}, C., {Frontera}, F., {Della Valle}, M., {Finelli}, F.,
  {Landi}, R.,  \& {Montanari}, E. 2008, \mnras, 391, 577

\bibitem[\protect\citeauthoryear{{Arnaud}}{{Arnaud}}{1996}]{ar1996}
{Arnaud}, K.~A. 1996, in ASP Conf. Ser. 101: Astronomical Data Analysis
  Software and Systems V, Vol.~5, 17

\bibitem[\protect\citeauthoryear{{Baumgardt} et~al.}{{Baumgardt}
  et~al.}{2006}]{2006MNRAS.372..467B}
{Baumgardt}, H., {Hopman}, C., {Portegies Zwart}, S.,  \& {Makino}, J. 2006,
  \mnras, 372, 467

\bibitem[\protect\citeauthoryear{{Belczynski} et~al.}{{Belczynski}
  et~al.}{2010}]{2010ApJ...714.1217B}
{Belczynski}, K., {Bulik}, T., {Fryer}, C.~L., {Ruiter}, A., {Valsecchi}, F.,
  {Vink}, J.~S.,  \& {Hurley}, J.~R. 2010, \apj, 714, 1217

\bibitem[\protect\citeauthoryear{{Blanton} et~al.}{{Blanton}
  et~al.}{2005}]{2005ApJ...631..208B}
{Blanton}, M.~R., {Lupton}, R.~H., {Schlegel}, D.~J., {Strauss}, M.~A.,
  {Brinkmann}, J., {Fukugita}, M.,  \& {Loveday}, J. 2005, \apj, 631, 208

\bibitem[\protect\citeauthoryear{{Blinnikov} et~al.}{{Blinnikov}
  et~al.}{1990}]{1990SvA....34..595B}
{Blinnikov}, S.~I., {Imshennik}, V.~S., {Nadezhin}, D.~K., {Novikov}, I.~D.,
  {Perevodchikova}, T.~V.,  \& {Polnarev}, A.~G. 1990, \sovast, 34, 595

\bibitem[\protect\citeauthoryear{{Bloom} et~al.}{{Bloom}
  et~al.}{2011}]{2011Sci...333..203B}
{Bloom}, J.~S., et~al. 2011, Science, 333, 203

\bibitem[\protect\citeauthoryear{{Campana} et~al.}{{Campana}
  et~al.}{2011}]{2011Natur.480...69C}
{Campana}, S., et~al. 2011, \nat, 480, 69

\bibitem[\protect\citeauthoryear{{Campana} et~al.}{{Campana}
  et~al.}{2006}]{2006Natur.442.1008C}
{Campana}, S., et~al. 2006, \nat, 442, 1008

\bibitem[\protect\citeauthoryear{{Cash}}{{Cash}}{1979}]{1979ApJ...228..939C}
{Cash}, W. 1979, \apj, 228, 939

\bibitem[\protect\citeauthoryear{{Colpi}, {Shapiro}, \& {Teukolsky}}{{Colpi}
  et~al.}{1993}]{1993ApJ...414..717C}
{Colpi}, M., {Shapiro}, S.~L.,  \& {Teukolsky}, S.~A. 1993, \apj, 414, 717

\bibitem[\protect\citeauthoryear{{de Vaucouleurs} et~al.}{{de Vaucouleurs}
  et~al.}{1991}]{1991rc3..book.....D}
{de Vaucouleurs}, G., {de Vaucouleurs}, A., {Corwin}, H.~G., Jr., {Buta},
  R.~J., {Paturel}, G.,  \& {Fouqu{\'e}}, P. 1991, {Third Reference Catalogue
  of Bright Galaxies. Volume I: Explanations and references. Volume II: Data
  for galaxies between 0$^{h}$ and 12$^{h}$. Volume III: Data for galaxies
  between 12$^{h}$ and 24$^{h}$.}

\bibitem[\protect\citeauthoryear{{Dickey} \& {Lockman}}{{Dickey} \&
  {Lockman}}{1990}]{1990ARA&A..28..215D}
{Dickey}, J.~M.,  \& {Lockman}, F.~J. 1990, \araa, 28, 215

\bibitem[\protect\citeauthoryear{{Esquej} et~al.}{{Esquej}
  et~al.}{2008}]{2008A&A...489..543E}
{Esquej}, P., et~al. 2008, \aap, 489, 543

\bibitem[\protect\citeauthoryear{{Fabian}}{{Fabian}}{2012}]{2012ARA&A..50..455F}
{Fabian}, A.~C. 2012, \araa, 50, 455

\bibitem[\protect\citeauthoryear{{Farrell} et~al.}{{Farrell}
  et~al.}{2009}]{2009Natur.460...73F}
{Farrell}, S.~A., {Webb}, N.~A., {Barret}, D., {Godet}, O.,  \& {Rodrigues},
  J.~M. 2009, \nat, 460, 73

\bibitem[\protect\citeauthoryear{{Fender} \& {Belloni}}{{Fender} \&
  {Belloni}}{2012}]{2012Sci...337..540F}
{Fender}, R.,  \& {Belloni}, T. 2012, Science, 337, 540

\bibitem[\protect\citeauthoryear{{Fruscione} et~al.}{{Fruscione}
  et~al.}{2006}]{2006SPIE.6270E..60F}
{Fruscione}, A., et~al. 2006, in Society of Photo-Optical Instrumentation
  Engineers (SPIE) Conference Series, Vol. 6270, Society of Photo-Optical
  Instrumentation Engineers (SPIE) Conference Series

\bibitem[\protect\citeauthoryear{{Fryer} et~al.}{{Fryer}
  et~al.}{2012}]{2012ApJ...749...91F}
{Fryer}, C.~L., {Belczynski}, K., {Wiktorowicz}, G., {Dominik}, M., {Kalogera},
  V.,  \& {Holz}, D.~E. 2012, \apj, 749, 91

\bibitem[\protect\citeauthoryear{{Garmire}}{{Garmire}}{1997}]{1997AAS...190.3404G}
{Garmire}, G.~P. 1997, in Bulletin of the American Astronomical Society,
  Vol.~29, American Astronomical Society Meeting Abstracts \#190, 823

\bibitem[\protect\citeauthoryear{{Gezari} et~al.}{{Gezari}
  et~al.}{2008}]{2008ApJ...676..944G}
{Gezari}, S., et~al. 2008, \apj, 676, 944

\bibitem[\protect\citeauthoryear{{Greene} \& {Ho}}{{Greene} \&
  {Ho}}{2007}]{2007ApJ...670...92G}
{Greene}, J.~E.,  \& {Ho}, L.~C. 2007, \apj, 670, 92

\bibitem[\protect\citeauthoryear{{Guillochon} \& {Ramirez-Ruiz}}{{Guillochon}
  \& {Ramirez-Ruiz}}{2013}]{2013ApJ...767...25G}
{Guillochon}, J.,  \& {Ramirez-Ruiz}, E. 2013, \apj, 767, 25

\bibitem[\protect\citeauthoryear{{Harris}}{{Harris}}{1996}]{1996AJ....112.1487H}
{Harris}, W.~E. 1996, \aj, 112, 1487

\bibitem[\protect\citeauthoryear{{Hillebrandt} \& {Niemeyer}}{{Hillebrandt} \&
  {Niemeyer}}{2000}]{2000ARA&A..38..191H}
{Hillebrandt}, W.,  \& {Niemeyer}, J.~C. 2000, \araa, 38, 191

\bibitem[\protect\citeauthoryear{{Jonker} et~al.}{{Jonker}
  et~al.}{2012}]{2012ApJ...758...28J}
{Jonker}, P.~G., et~al. 2012, \apj, 758, 28

\bibitem[\protect\citeauthoryear{{Kasen}}{{Kasen}}{2010}]{2010ApJ...708.1025K}
{Kasen}, D. 2010, \apj, 708, 1025

\bibitem[\protect\citeauthoryear{{Komossa} \& {Bade}}{{Komossa} \&
  {Bade}}{1999}]{1999A&A...343..775K}
{Komossa}, S.,  \& {Bade}, N. 1999, \aap, 343, 775

\bibitem[\protect\citeauthoryear{{Komossa} et~al.}{{Komossa}
  et~al.}{2004}]{2004ApJ...603L..17K}
{Komossa}, S., {Halpern}, J., {Schartel}, N., {Hasinger}, G., {Santos-Lleo},
  M.,  \& {Predehl}, P. 2004, \apjl, 603, L17

\bibitem[\protect\citeauthoryear{{Krolik} \& {Piran}}{{Krolik} \&
  {Piran}}{2011}]{2011ApJ...743..134K}
{Krolik}, J.~H.,  \& {Piran}, T. 2011, \apj, 743, 134

\bibitem[\protect\citeauthoryear{{Lawrence} et~al.}{{Lawrence}
  et~al.}{2007}]{2007MNRAS.379.1599L}
{Lawrence}, A., et~al. 2007, \mnras, 379, 1599

\bibitem[\protect\citeauthoryear{{Levan} et~al.}{{Levan}
  et~al.}{2011}]{2011Sci...333..199L}
{Levan}, A.~J., et~al. 2011, Science, 333, 199

\bibitem[\protect\citeauthoryear{{Lodato} \& {Rossi}}{{Lodato} \&
  {Rossi}}{2011}]{2011MNRAS.410..359L}
{Lodato}, G.,  \& {Rossi}, E.~M. 2011, \mnras, 410, 359

\bibitem[\protect\citeauthoryear{{Luminet} \& {Pichon}}{{Luminet} \&
  {Pichon}}{1989}]{1989A&A...209..103L}
{Luminet}, J.-P.,  \& {Pichon}, B. 1989, \aap, 209, 103

\bibitem[\protect\citeauthoryear{{Madau} \& {Rees}}{{Madau} \&
  {Rees}}{2001}]{2001ApJ...551L..27M}
{Madau}, P.,  \& {Rees}, M.~J. 2001, \apjl, 551, L27

\bibitem[\protect\citeauthoryear{{M{\'e}sz{\'a}ros}}{{M{\'e}sz{\'a}ros}}{2006}]{2006RPPh...69.2259M}
{M{\'e}sz{\'a}ros}, P. 2006, Reports on Progress in Physics, 69, 2259

\bibitem[\protect\citeauthoryear{{Miller} \& {Hamilton}}{{Miller} \&
  {Hamilton}}{2002}]{2002MNRAS.330..232C}
{Miller}, M.~C.,  \& {Hamilton}, D.~P. 2002, \mnras, 330, 232

\bibitem[\protect\citeauthoryear{{Nakar} \& {Piran}}{{Nakar} \&
  {Piran}}{2011}]{2011Natur.478...82N}
{Nakar}, E.,  \& {Piran}, T. 2011, \nat, 478, 82

\bibitem[\protect\citeauthoryear{{Orosz} et~al.}{{Orosz}
  et~al.}{2007}]{2007Natur.449..872O}
{Orosz}, J.~A., et~al. 2007, \nat, 449, 872

\bibitem[\protect\citeauthoryear{{{\"O}zel} et~al.}{{{\"O}zel}
  et~al.}{2010}]{2010ApJ...725.1918O}
{{\"O}zel}, F., {Psaltis}, D., {Narayan}, R.,  \& {McClintock}, J.~E. 2010,
  \apj, 725, 1918

\bibitem[\protect\citeauthoryear{{Page} et~al.}{{Page}
  et~al.}{2004}]{2004ApJS..155..623P}
{Page}, D., {Lattimer}, J.~M., {Prakash}, M.,  \& {Steiner}, A.~W. 2004, \apjs,
  155, 623

\bibitem[\protect\citeauthoryear{{Portegies Zwart} \& {McMillan}}{{Portegies
  Zwart} \& {McMillan}}{2002}]{2002ApJ...576..899P}
{Portegies Zwart}, S.~F.,  \& {McMillan}, S.~L.~W. 2002, \apj, 576, 899

\bibitem[\protect\citeauthoryear{{Ramirez-Ruiz} \& {Rosswog}}{{Ramirez-Ruiz} \&
  {Rosswog}}{2009}]{2009ApJ...697L..77R}
{Ramirez-Ruiz}, E.,  \& {Rosswog}, S. 2009, \apjl, 697, L77

\bibitem[\protect\citeauthoryear{{Randall} et~al.}{{Randall}
  et~al.}{2008}]{2008ApJ...688..208R}
{Randall}, S., {Nulsen}, P., {Forman}, W.~R., {Jones}, C., {Machacek}, M.,
  {Murray}, S.~S.,  \& {Maughan}, B. 2008, \apj, 688, 208

\bibitem[\protect\citeauthoryear{{Rezzolla} et~al.}{{Rezzolla}
  et~al.}{2011}]{2011ApJ...732L...6R}
{Rezzolla}, L., {Giacomazzo}, B., {Baiotti}, L., {Granot}, J., {Kouveliotou},
  C.,  \& {Aloy}, M.~A. 2011, \apjl, 732, L6

\bibitem[\protect\citeauthoryear{{Rosswog}, {Ramirez-Ruiz}, \& {Hix}}{{Rosswog}
  et~al.}{2009}]{2009ApJ...695..404R}
{Rosswog}, S., {Ramirez-Ruiz}, E.,  \& {Hix}, W.~R. 2009, \apj, 695, 404

\bibitem[\protect\citeauthoryear{{Sakamoto} et~al.}{{Sakamoto}
  et~al.}{2005}]{2005ApJ...629..311S}
{Sakamoto}, T., et~al. 2005, \apj, 629, 311

\bibitem[\protect\citeauthoryear{{Shcherbakov} et~al.}{{Shcherbakov}
  et~al.}{2012}]{2012arXiv1212.4837S}
{Shcherbakov}, R.~V., {Pe'er}, A., {Reynolds}, C.~S., {Haas}, R., {Bode}, T.,
  \& {Laguna}, P. 2012, ArXiv e-prints

\bibitem[\protect\citeauthoryear{{Stritzinger} \& {Sollerman}}{{Stritzinger} \&
  {Sollerman}}{2007}]{2007A&A...470L...1S}
{Stritzinger}, M.,  \& {Sollerman}, J. 2007, \aap, 470, L1

\bibitem[\protect\citeauthoryear{{Strubbe} \& {Quataert}}{{Strubbe} \&
  {Quataert}}{2009}]{2009MNRAS.400.2070S}
{Strubbe}, L.~E.,  \& {Quataert}, E. 2009, \mnras, 400, 2070

\bibitem[\protect\citeauthoryear{{van Eerten} \& {MacFadyen}}{{van Eerten} \&
  {MacFadyen}}{2011}]{2011ApJ...733L..37V}
{van Eerten}, H.~J.,  \& {MacFadyen}, A.~I. 2011, \apjl, 733, L37

\bibitem[\protect\citeauthoryear{{Volonteri}}{{Volonteri}}{2010}]{2010Natur.466.1049V}
{Volonteri}, M. 2010, \nat, 466, 1049

\bibitem[\protect\citeauthoryear{{Volonteri}}{{Volonteri}}{2012}]{2012Sci...337..544V}
{Volonteri}, M. 2012, Science, 337, 544

\bibitem[\protect\citeauthoryear{{Wang} \& {Merritt}}{{Wang} \&
  {Merritt}}{2004}]{2004ApJ...600..149W}
{Wang}, J.,  \& {Merritt}, D. 2004, \apj, 600, 149

\end{thebibliography}

\section*{Acknowledgments} \noindent PGJ acknowledges the hospitality
of the Institute of Astronomy, Cambridge where much of this work was
done and PGJ also acknowledges discussions with Elena Rossi and Sjoerd
van Velzen. Based on observations obtained with MegaPrime/MegaCam, a
joint project of CFHT and CEA/IRFU, at the Canada-France-Hawaii
Telescope (CFHT) which is operated by the National Research Council
(NRC) of Canada, the Institut National des Science de l'Univers of the
Centre National de la Recherche Scientifique (CNRS) of France, and the
University of Hawaii. This work is based in part on data products
produced at Terapix available at the Canadian Astronomy Data Centre as
part of the Canada-France-Hawaii Telescope Legacy Survey, a
collaborative project of NRC and CNRS. Based on observations obtained with MegaPrime/MegaCam, a joint
  project of CFHT and CEA/IRFU, at the Canada-France-Hawaii Telescope
  (CFHT) which is operated by the National Research Council (NRC) of
  Canada, the Institut National des Science de l'Univers of the Centre
  National de la Recherche Scientifique (CNRS) of France, and the
  University of Hawaii. This work is based in part on data products
  produced at Terapix available at the Canadian Astronomy Data Centre
  as part of the Canada-France-Hawaii Telescope Legacy Survey, a
  collaborative project of NRC and CNRS.  

  Funding for SDSS-III has been provided by the Alfred P. Sloan
  Foundation, the Participating Institutions, the National Science
  Foundation, and the U.S. Department of Energy Office of Science. The
  SDSS-III web site is http://www.sdss3.org/.

  SDSS-III is managed by the Astrophysical Research Consortium for the
  Participating Institutions of the SDSS-III Collaboration including
  the University of Arizona, the Brazilian Participation Group,
  Brookhaven National Laboratory, University of Cambridge, Carnegie
  Mellon University, University of Florida, the French Participation
  Group, the German Participation Group, Harvard University, the
  Instituto de Astrofisica de Canarias, the Michigan State/Notre
  Dame/JINA Participation Group, Johns Hopkins University, Lawrence
  Berkeley National Laboratory, Max Planck Institute for Astrophysics,
  Max Planck Institute for Extraterrestrial Physics, New Mexico State
  University, New York University, Ohio State University, Pennsylvania
  State University, University of Portsmouth, Princeton University,
  the Spanish Participation Group, University of Tokyo, University of
  Utah, Vanderbilt University, University of Virginia, University of
  Washington, and Yale University.

\end{document}